\documentclass[12pt]{iopart}

\usepackage{iopams}

\expandafter\let\csname equation*\endcsname\relax
\expandafter\let\csname endequation*\endcsname\relax

\usepackage{amsmath}
\usepackage{color,soul}
\usepackage{array,graphicx,caption}
\usepackage{braket}

\begin{document}

\title[Speeding up qubit control with bipolar single-flux-quantum pulse sequences]{Speeding up qubit control with bipolar single-flux-quantum pulse sequences}

\author{Vsevolod Vozhakov$^{1,2,3}$, Marina Bastrakova$^{2}$, Nikolay Klenov$^{2,4}$, Arkady Satanin$^{5}$, Igor Soloviev$^{1,2}$}
\address{$^1$Lomonosov Moscow State University Skobeltsyn Institute of Nuclear Physics, 119991 Moscow, Russia}
\address{$^2$Lobachevsky State University of Nizhny Novgorod, Nizhny Novgorod 603950, Russia}
\address{$^3$Russian Quantum Center, 143025 Skolkovo, Moscow, Russia}
\address{$^4$Faculty of Physics, Lomonosov Moscow State University, 119991 Moscow, Russia}
\address{$^5$Russia National Research University Higher School of Economics, Moscow 101000, Russia}

\ead{igor.soloviev@gmail.com}

\vspace{10pt}

\begin{abstract}
The development of quantum computers based on superconductors requires the improvement of the qubit state control approach aimed at the increase of the hardware energy efficiency. A promising solution to this problem is the use of superconducting digital circuits operating with single-flux-quantum (SFQ) pulses, moving the qubit control system into the cold chamber. However, the qubit gate time under SFQ control is still longer than under conventional microwave driving. Here we introduce the bipolar SFQ pulse control based on ternary pulse sequences. We also develop a robust optimization algorithm for finding a sequence structure that minimizes the leakage of the transmon qubit state from the computational subspace. We show that the appropriate sequence can be found for arbitrary system parameters from the practical range. The proposed bipolar SFQ control reduces a single qubit gate time by halve compared to nowadays unipolar SFQ technique, while maintaining the gate fidelity over 99.99\%.
\end{abstract}

\section{Introduction}
Quantum computing is a fast developing frontier technology. Superconducting materials are shown to be one of the most promising physical basis for it's implementation \cite{QRevDS}. A number of superconducting quantum processors containing a few dozens of qubits are realized nowadays \cite{Vozhakov, google, fluxonium}. It is expected that their further advance will allow tackling the computationally complex problems arising in the areas of 
quantum chemistry \cite{qchemistry}, many-body system simulations \cite{randall2021}, quantum cryptography \cite{cryptography}, deep learning \cite{deeplearning} etc.

The workhorse of quantum processors is a charge qubit of a transmon type \cite{Koch2007}. It can be thought of as an LC-oscillator where the inductor is substituted for a Josephson junction \cite{Krantz2019}. The nonlinear Josephson inductance introduces slight anharmonicity providing non-equidistant distribution of the energy levels. The latter allows one to separate a subspace of two levels for computation purposes. The predominance of the Josephson energy over the energy stored in the capacitance makes the qubit almost insensitive to the charge noise. Unfortunately, this fact also limits anharmonicity and minimum control pulse duration \cite{Koch2007}.

In a standard microwave technique, the control pulse is generated using quadrature mixer (QM) by combining the high-frequency signal from a room-temperature local oscillator (LO) with the pulse envelope provided by an  arbitrary waveform generator (AWG). The resulting drive pulse is typically resonant with the qubit. For a typical qubit frequency corresponding to the transition between between the ground  $\left|0\right>$ and first excited state $\left|1\right>$, $\omega_{01}/2\pi \sim 5$~GHz, the control pulse duration providing a $\pi$ rotation on the Bloch sphere (states population inversion) with more than 99.99\% fidelity is typically above 40~ns \cite{Gambetta2011}. By using the derivative removal by adiabatic gate (DRAG) strategy proposed in \cite{DRAG}, this time can be approximately halved \cite{DRAG1, DRAG2, DRAG3}.

While quantum supremacy is already demonstrated on a notional task \cite{google}, and quantum correction becoming possible \cite{andersen2020, egan2021, Krinner2022}, still the fault-tolerant quantum computing requires a massive hardware overhead: systems with up to millions of physical qubits is far beyond current capabilities.
Since maintenance of each qubit requires a number of room-temperature devices (LO, AWG, QM, etc.), this leads to a prohibitively complex and costly system. Many cables and connectors coupling room-temperature hardware with a cold chamber provide heat inflow channels \cite{Krinner2019} increasing qubit relaxation rates and causing related control signal distortion (see Supplementary materials of \cite{neill2018}). The long latency of the feedback loop associated with the signal traveling between the room-temperature devices and cold chamber limits the system performance \cite{Vijay2012, Salathe2018}. 

There two main approaches to circumvent these issues. The first is the use of frequency division multiplexing in the microwave domain \cite{Huang2022} or utilization of photonic links with optional wavelength division multiplexing \cite{optic2021,optic2_2021,Joshi2022}. This allows the decrease of the heat inflow associated with cables at the cost of more sophisticated transmission of microwave signals. However, the amount of room temperature electronics here is still proportional to the number of qubits, while the delays associated with the distance between the room temperature controllers and the quantum system remain large.

The second approach is the miniaturization of some room-temperature hardware and moving it closer to the quantum chip, e.g., to 3 or 4 K temperature stage of the cryocooler. Cryogenic CMOS microwave sources and digital-to-analog converters are being developed in this line of research \cite{ColdCMOS,CryoCMOS2020,Xue2021,CryoCMOS2021}. In front of advantages of the maturity of CMOS technology, the disadvantages of cryo-CMOS control are relatively high power consumption and thermal noise. Another promising candidate here is the superconducting digital electronics known for its high energy efficiency combined with high clock rates \cite{Beil, IRDS2022}. In these digital circuits, a bit of information is represented by a short unipolar voltage pulse that occurs when a single flux quantum (SFQ) passes through a Josephson junction. Irradiation of a qubit with an SFQ pulse leads to a slight change in the qubit state. Successive application of a bitstream provides the state vector controlled rotation \cite{QCIMukh}. Interestingly, the bitstream structure can be optimized to drive the qubits with different transition frequencies at the fixed digital circuit clock \cite{McD1}. Therefore, this approach no longer needs LO, AWG and QM for each qubit. 

Due to the wide spectrum of short picosecond SFQ pulse and the small anharmonicity of transmons, the pulse amplitude cannot be large in order to avoid leakage of the qubit state from the computational subspace. In contrast to the flux qubit \cite{Bastrakova2019, Bastrakova2020, Bastrakova2022}, in the case of using a simple regular SFQ pulse sequence to control the transmon, the length of the sequence should be sufficiently large.
If the pulse repetition rate is synchronized with the transmon qubit frequency, a $\pi/2$ rotation with 99.99~\% fidelity can be obtained with a regular sequence containing $\sim 300$~pulses \cite{McD0}. This turns into $\sim 60$~ns gate time or over 100~ns for a $\pi$ rotation.

By using the so-called scalable leakage optimized pulse sequences (SCALLOPS) \cite{McD1} utilizing more than one pulse during the qubit oscillation period, $2\pi/\omega_{01}$, the gate time can be drastically improved. The work \cite{McD1} presents a $\pi/2$ rotation for qubits with frequencies about $\omega_{01}/2\pi \approx 5$~GHz at gate time above $12$~ns. Therefore, one can expect the gate time similar to the one obtained in the frame of the conventional microwave DRAG technique for a $\pi$ rotation.

This paper further develops the SCALLOP approach by introducing the pulses with negative polarity into the control sequences. After consideration of possible implementation of bipolar SFQ control, we optimize the bipolar SFQ pulse sequences presenting an algorithm aimed at the minimization of leakage from the computational subspace. Finally, we show that the proposed modernization of SCALLOP allows one to additionally reduce a single qubit gate time by half.

\section{SFQ control}

There are two ways to get a regular SFQ pulse signal. One can use a DC-to-SFQ converter \cite{likharev1991,McDer2019} to transform the reference microwave signal from LO to an SFQ pulse train (one SFQ is generated per signal period). Alternatively, the pulse train can be obtained by using an on-chip long annular Josephson junction clock source, with successive digital frequency downscaling if needed \cite{Kirichenko2005}. This pulse signal can be directly utilized to drive a transmon for its simple characterization \cite{McDer2019,Howe2022}.

The pulse train can be further processed by digital circuits. For example, an SFQ switch may be used to form the pulse sequence of a certain length \cite{He2022}. The pulses then can be digitally multiplied to increase their amplitude \cite{Sol2005,Febvre2022}. The obtained waveform may be optionally passed through a bandpass filter to get a microwave driving pulse quite similar to the standard one \cite{Febvre2022,He2020}.

\begin{figure}[b]
\center{\includegraphics[scale=0.25]{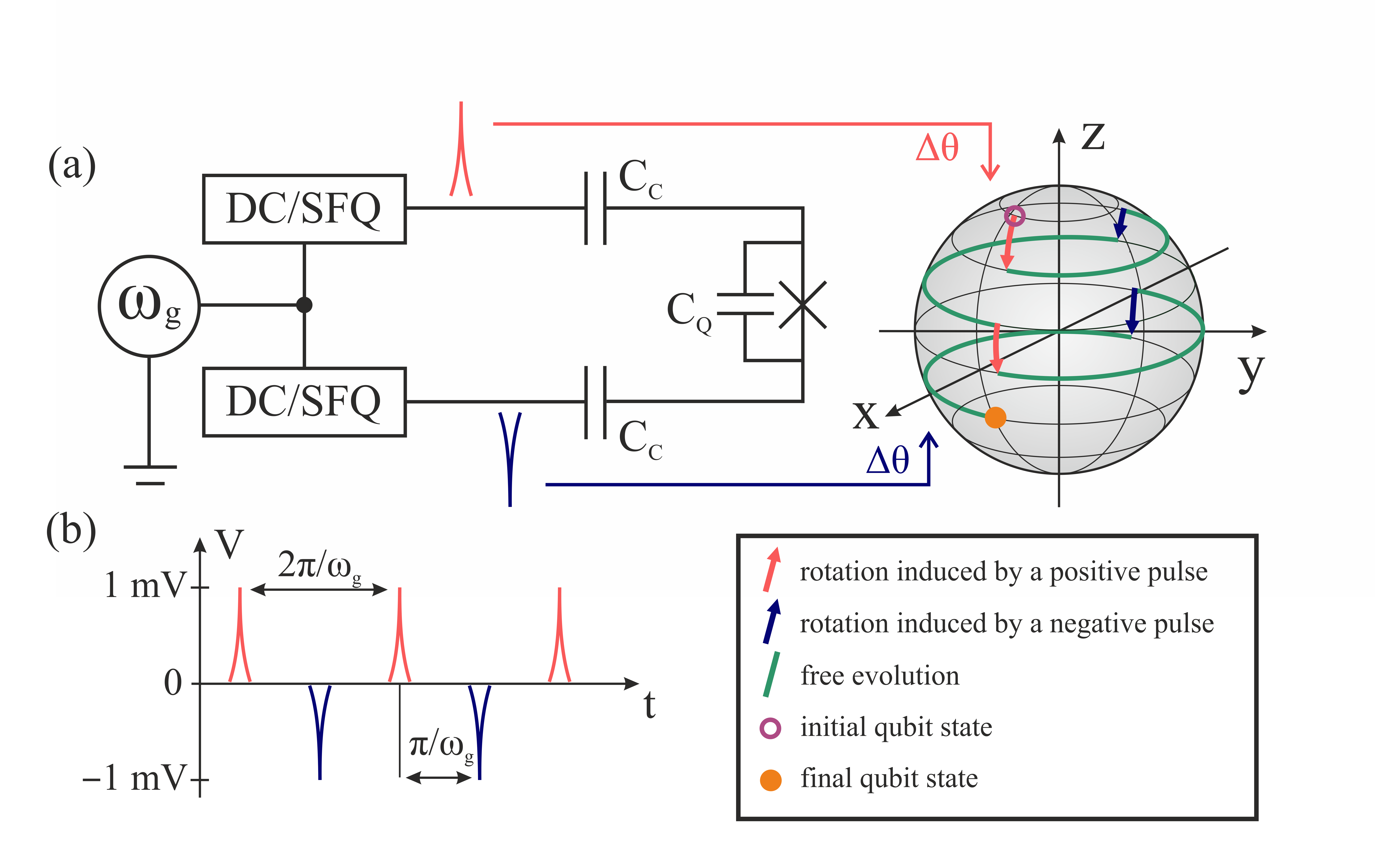}}
\caption{(a) Bipolar SFQ pulse generator composed of two DC-to-SFQ converters connected to a transmon. The qubit state changing under effect of a regular bipolar SFQ pulse sequence is illustrated on a Bloch sphere. (b) Schematic representation of a regular bipolar pulse sequence on the time axis.}
\label{josring:image}
\end{figure}

The utilization of a more complex SFQ co-processor allows operating with even more sophisticated form of the driving signals. The pulse sequence of an arbitrary structure can be generated with a pulse generator at room temperature and then pre-loaded into a superconducting register file at a low frequency, with voltage pulses converted to SFQ by a DC-to-SFQ converter. The stored sequences then can be repeatedly utilized in the execution of different quantum operations \cite{QCIMukh} at high clock frequency of the on-chip clock generator. By complementing this control with a measurement system also based on the superconducting circuits \cite{Opremcak2018,Howington2019}, including superconducting LO \cite{Yan2021}, one can design a fully cryogenic quantum–classical interface.

In all considered cases, the signal is based on a bitstream, where a bit means one or zero physically represented as a short SFQ voltage pulse or its absence on the clock period, $T_g = 2\pi/\omega_g$, where $\omega_g$ is the clock generator frequency; an SFQ pulse is of a picosecond width, $\tau_p \ll T_g$ whereas its voltage integral over time is equal to a single flux quantum, $\displaystyle \Phi_0 = \int^{\tau_p}_{0}V(t)dt$. An intuitive idea of an appropriate control pulse sequence structure can be obtained from analogy with Ramsey interference, showing the most efficient level population inversion if the microwave pulses are applied with the delay $2\pi k/\omega_{01}$ (where $k$ is integer). Similarly, the simplest SFQ drive signal can be designed as a pulse train (the bitstream composed of ones only) synchronized with the qubit frequency, $\omega_g = \omega_{01}$. However, if the clock frequency is twice the qubit frequency, the bitstream should be ones interlaced with zeros: $1,0,1,0,\ldots$. Otherwise, every second pulse will compensate for the impact of every first one, how does it happen in the case where Ramsey pulses are applied with the delay $\pi/\omega_{01}$. In the case of arbitrary frequency excess, $\omega_g > \omega_{01}$, one can suggest putting ones in the bitstream in time windows $[-\pi/2 + 2\pi k,\pi/2 + 2\pi k]/\omega_{01}$ and zeros otherwise. 

It follows from the above that the pauses in the control pulse sequences (zeros in the bitstream) are mainly due to the unipolar nature of the signal. One could utilize the downtimes by moving to the ternary logic, in which a trit (a unit of ternary data) means $1$, $-1$ or $0$ physically represented as an SFQ voltage pulse of positive or negative polarity, or its absence on the clock period. For example, in the considered case, $\omega_g = 2\omega_{01}$, one could design the tritstream of ones alternating in sign: $1,-1,1,-1,\ldots$. The impact of pulses of both polarities is codirectional here, as in the case where the second Ramsey pulse, applied with a delay of $\pi/\omega_{01}$, has the opposite polarity with respect to the first one. 

The considered situation of the bipolar drive signal can be simply implemented by differential connection of two DC-to-SFQ converters driven with a shift of half a clock cycle, see figure~\ref{josring:image}. Assuming the utilization of a more complex SFQ co-processor, \cite{QCIMukh} where a tritstream can be hold as two synchronized bit sequences in a standard register file, or as a trit sequence in multibit storage cells \cite{Katam2020,Soloviev2021}, we further consider this particular case of tritstreams application to a qubit.


The trit sequences are composed of a large number of short in time and low in amplitude SFQ pulses. An SFQ pulse spectrum is broad and is nearly constant in the qubit frequency range. Therefore, the sequence spectrum is determined solely by the sequence structure, while pulse certain shape is not important \cite{McD0} and can be safely substituted for a rectangular one in the simulation (see also Supplementary materials for an example).

\section{Transmon under SFQ drive}

The transmon design is derived from a Cooper pair box with shunt capacitance $C_{Q}$ and the capacitance coupling voltage drive, $C_{C}$. The capacitor energy  $E_C = e^{2}/2C$ (where $C = C_{C} + C_{Q}$) is significantly less than the Josephson energy $E_J$ which also can be \textit{in situ} adjusted by applying an external magnetic flux to the circuit. This feature, $E_{C} \ll E_{J}$, makes transmons less susceptible to charge noise due to the decrease in the charge dispersion. The transmon Hamiltonian \cite{Koch2007} including the electrostatic and Josephson components can be written in the following form:
\begin{equation}
 \hat{H}_{q}=\frac{\left( \hat{Q} - C_{C}V(t) \right) ^{2}}{2C}-E_{J}\left( 1 - \cos \hat{\phi} \right). \label{eq:hamiltonian}
\end{equation}
It is convenient to express the charge operator, $\hat{Q} = 2e\hat{n}$, as Cooper pairs number operator $ \displaystyle \hat{n} = -i\frac{\partial}{\partial \phi}$ in the units of charge, $2e$, where $\hat{\phi}$ is the phase operator. In this case, $\left[ \hat{\phi}, \hat{n}\right] = i$. By making the variable substitutions, we can rewrite the transmon Hamiltonian in the form:
\begin{equation}
 \hat{H}_{q}=4E_{C} \left( \hat{n} - n_g \right)^{2}-E_{J}\left( 1 - \cos \hat{\phi} \right), \label{eq:hamiltonian2}
\end{equation}
where $n_g = C_{c}V(t)/2e$ is called the effective offset charge.

The Schrodinger equation with a Hamiltonian (\ref{eq:hamiltonian2}) can be solved exactly in terms of Mathieu functions \cite{Koch2007}. However, under the condition $E_{C} \ll E_{J}$, we assume that the phase is highly localized and it is possible to use the Hamiltonian of a weakly nonlinear oscillator with the expansion of $\cos \hat{\phi} = 1-\frac{\hat{\phi}^2}{2!}+\frac{\hat{\phi}^4}{4!} + ...$ using the fourth order Taylor series for approximation. 

We can approach the quantization, where we define the creation $\hat{a}^\dagger$ and annihilation  $\hat{a}$ operators in terms of the charge and phase fluctuations 
\begin{equation}
\begin{array}{cc}
     \hat{n}=-\frac{i}{2}\left( \frac{E_{J}}{2E_{C}}\right)^{1/4} \left(\hat{a}-\hat{a}^\dagger\right), \quad
     \hat{\phi}=\left(\frac{2E_{C}}{E_{J}}\right)^{1/4}\left(\hat{a}+\hat{a}^\dagger\right).
\end{array}\label{eq:operators}
\end{equation}
 Substituting expressions~(\ref{eq:operators}) into Eq.~(\ref{eq:hamiltonian2})  and neglecting constants that have no influence on transmon dynamics, we have the qubit Hamiltonian as (we put Planck's constant $\hbar = 1$):
\begin{equation}
    \hat{H}_{q} = \omega_{01}\hat{a}^\dagger\hat{a} + \mu \left(\hat{a}^\dagger + \hat{a}\right)^{4}+i \varepsilon (t)\left(\hat{a}-\hat{a}^\dagger\right), \label{eq:hamiltonian3}
\end{equation}
where qubit frequency can be calculated as $\displaystyle \omega_{01}=\sqrt{8E_{J}E_{C}}$, $ \displaystyle \mu = -E_C / 12$ is the parameter of nonlinearity and $\displaystyle \varepsilon(t)= \frac{C_{C}V(t)}{2} \sqrt{\frac{\omega_{01}}{2C_Q}}$ is the amplitude of the driving field. A single SFQ pulse induces a discrete small rotation by the angle $\displaystyle \Delta \theta = C_{C}\Phi_0 \sqrt{\frac{\omega_{01}}{2C_Q}} $ on the Bloch sphere corresponding to the qubit state change (see figure~\ref{josring:image}). The bipolar sequence consisting of $M$ SFQ pulses implements quantum gate $U_{g}$ by rotating the qubit state vector.  We calculate the evolution of the state populations by solving the time-dependent Schrodinger equation:
\begin{equation}
    \ket{\psi (t)} = U(t) \ket{\alpha}, \quad U(t) = \hat{P}e^{-i \int^{t}_{0} \hat{H}_{q}(\tau)d\tau},
\end{equation}
where $\hat{P}$ denotes the time-ordering operator, $\ket{\alpha}$ is the initial state along the cardinal directions of the Bloch sphere:
\begin{equation}
    \begin{array}{c}
     \ket{x_{\pm}} = \dfrac{\ket{0} \pm\ket{1}}{\sqrt{2}}, \quad
        \ket{y_{\pm}} = \dfrac{\ket{0}\pm i \ket{1}}{\sqrt{2}}, \quad
        \ket{z_{+}} = \ket{0}, \quad \ket{z_{-}}  = \ket{1}. 
    \end{array} \label{eq5}
\end{equation}
The level population dynamics is defined as
\begin{equation}
    W_{m}(t) = \left| \left<m|\psi(t)\right>\right|^{2}, \quad  m = 0, 1, 2... \label{eq6}
\end{equation}
The averaged fidelity \cite{Bowdrey} over all initial states is
\begin{equation}
    \langle F \rangle = \frac{1}{6}\sum_{|\alpha \rangle} \left| \left<\alpha|U^{\dagger}U_{g}|\alpha\right>\right|^{2},
\end{equation}
where the summation runs over the states \eqref{eq5} and the matrix of the operator $U_g$ corresponds to an ideal gate in a qubit subspace.


\section{Bipolar modification of SCALLOP sequences}

In the SCALLOPS approach \cite{McD1}, the control sequence is composed of several identical subsequences, where the unipolar SFQ pulses are placed symmetrically in time in the chosen half cycle of qubit precession around the Bloch sphere. Such pulse arrangement provides a co-directional rotation of qubit state vector around the corresponding axis (X or Y). 


However, there are no restrictions on the utilization of the entire qubit precession time when using SFQ pulses of both positive and negative polarity, see figure~\ref{josring:image}. By doing that, we can increase the pulse density in a sequence about twice. 
Intuitively, it seems that such an approach can lead to a reduction in the gate time by half. Since this time is far less than dephasing and relaxation ones, the main source of error is the leakage to the higher qubit states, $m \geq 2$. In the next Section, we present an algorithm that optimizes an SFQ pulse sequence structure to minimize this leakage, and accordingly, the gate infidelity. Note, that contrary to the work \cite{McD1}, we do not restrict the pulse placement to symmetrical pairs. We also assume a ``virtual Z-gate'' \cite{zgate} after the pulse sequence implementation in order to reach the desired wave function phase (see Supplementary materials). Besides, the introduction of a negative pulse polarity further expands the search space.

\section{Bipolar sequence optimization algorithm}

\begin{figure}[t]
\center{\includegraphics[scale=0.45]{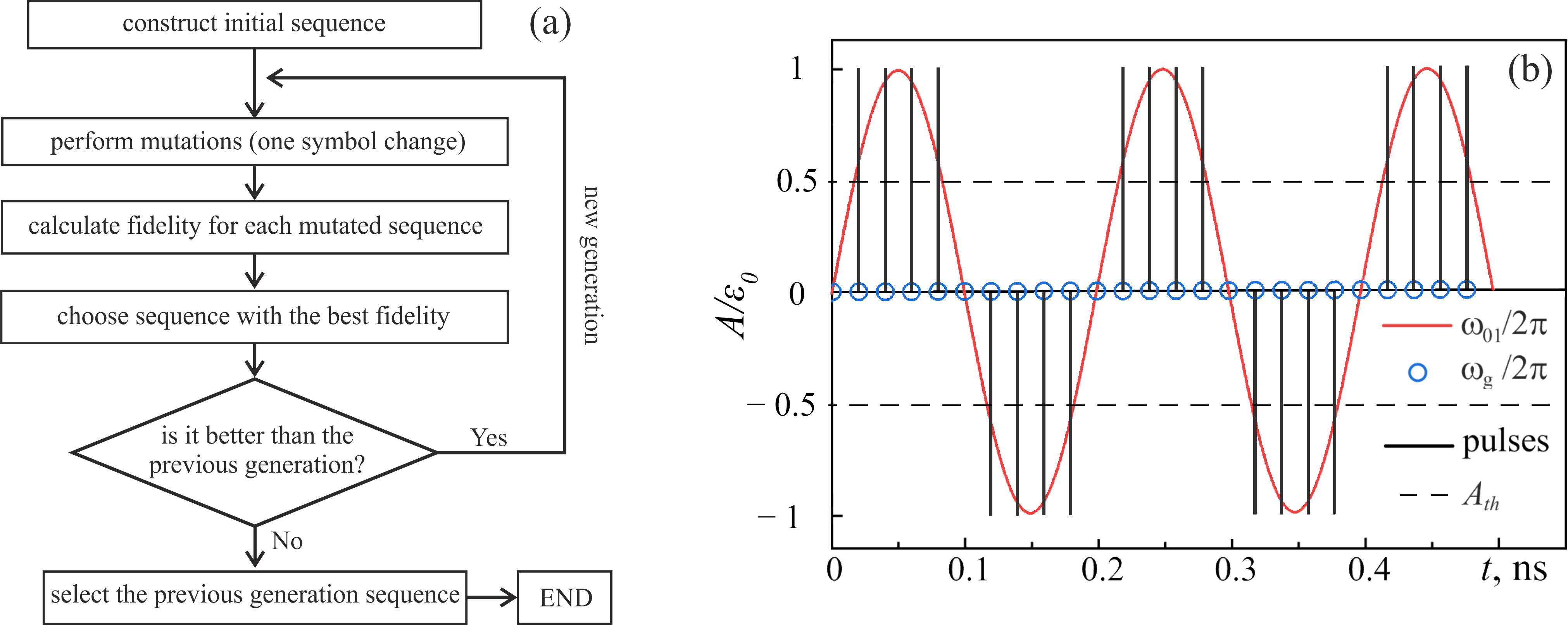}}
\caption{(a) Optimization algorithm flowchart. (b) Initial sequence with $A_{th} = 0.5 \varepsilon_0$, where harmonic signal amplitude $\varepsilon_0 = 0.052$~GHz; qubit frequency $\omega_{01}/2\pi = 5$~GHz and clock generator frequency $\omega_g/2\pi = 50$~GHz. The blue circles show periods of the generator where pulses can be placed. If the harmonic qubit frequency signal (red line) exceeds the threshold (black dash line), a pulse of the corresponding polarity is placed (black solid line).}

\label{algscheme:image}
\end{figure}


We have developed a robust optimization algorithm to find a control sequence suitable for a transmon with arbitrary parameters from a practical range: $\omega_{01}/2\pi \in [3...7]$~GHz, $\mu/2\pi \in [0.2...0.45]$~GHz, with the drive coupling corresponding to a single pulse rotation angle in the range $\Delta\theta \in [0.021...0.033]$~rad, and the clock generator frequency, $\omega_g/2\pi \in [25...50]$~GHz. We consider a single qubit gate on the example of a $\pi/2$ rotation around $Y$ axis, $U_g = Y_{\pi/2}$. The framework of the algorithm is close to a discrete coordinate descent. It includes the following steps (see figure~\ref{algscheme:image}a).  



\begin{enumerate}
    \item 
    \textit{Creating an initial sequence of the length $M$}. 
    We take a harmonic signal on the qubit frequency, $\omega_{01}$, with the amplitude of the exciting field, $\varepsilon_0$. Then we choose a certain threshold value $A_{th} \leq \varepsilon_0$, e.g., $A_{th} = 0.5 \varepsilon_0$. We place SFQ pulses into the clock generator periods on the generator frequency, $\omega_{g}$, when the harmonic signal exceeds the threshold, see figure~\ref{algscheme:image}b. The pulse polarity is defined by the sign of the harmonic signal.
    
    \item 
    \textit{Sequence mutations.} 
    
    Since the considered trit sequence is composed of ``$1$", ``$-1$" or ``$0$" symbols, we define a single mutation as a change of one symbol to one of the two remaining options. In this way, we start every generation by constructing a set of all possible mutated sequences (one initial sequence produces $2M$ mutated sequences overall). For each sequence in the set we count it's fidelity, $\langle F \rangle$, and the optimal single pulse rotation angle, $\Delta\theta$.  
    Finally, we choose the best sequence in accordance with the fidelity value as a starting sequence for the next generation. Alternatively, it is possible to choose a set of good initial sequences instead of a single best one to expand the search landscape. The choice of a particular option at this stage depends on the convergence of the algorithm for a given set of search parameters.
    \item
    \textit{Evolution.} The evolution process continues until $ \langle F_i \rangle < \langle F_{i-1} \rangle $, where $i$ is the generation number. 
    After the process termination, we check the single pulse rotation angle of the final sequence. If it is close enough to the desired value (the angle precision was set to $10^{-4}$ rad), the optimal control sequence is found.
    \item
    \textit{Optimal single pulse rotation angle search.} In case of angle mismatch, we change the length of the sequence. We increase or decrease the sequence length $M$ by 1 if $\Delta \theta_{found} > \Delta \theta_{desired}$ or vice versa, accordingly. A new initial sequence for the new $M$ is created as described in step (i) followed by repetition of the evolution process. We keep changing the length of the sequence until a sequence with the desired angle is found.
\end{enumerate}

If the algorithm has not found any sequences fit with the desired criterion, it is possible to change the optimal threshold for starting sequence $A^{opt}_{th}$ or doing some pre-calculations with different threshold values to estimate the optimal starting point. For example, it is possible to make an estimation based on a linear interpolation.  In order to do that, we perform the evolution process and calculate the angles $\Delta\theta_{min}$ and $\Delta\theta_{max}$ for initial sequences constructed with the minimum, $ A_{min} = 0.01 \varepsilon_0$, and maximum, $A_{max} = \varepsilon_0$, threshold values. The desired value is then found using linear interpolation, $A^{opt}_{th} = \Delta\theta \frac{A_{max}-A_{min}}{\Delta\theta_{max}-\Delta\theta_{min}}-A_{min}$. If the desired sequence still cannot be found, some sequences providing good fidelity found earlier in the mutation process can be used as new search seeds instead of those constructed in step (i).

\begin{table*}[h]
\caption{Control sequences providing $Y_{\pi/2}$ gate obtained using the optimization algorithm for various transmon parameters under a bipolar SFQ drive.}
\centering
\begin{tabular}{|c|c|c|c|c|c|c|c|}
    \hline
    N & $\Delta\theta$,~rad & $\omega_{01}/2\pi $,~GHz & $\mu /2\pi $,~GHz & $\omega_{g}/2\pi $,~GHz & $1 - \langle F \rangle$, $10^{-5}$ & Length & $t_{\pi/2}$,~ns \\
    \hline
    1 & 0.024 & 3 & 0.25 & 25 & 6.729 & 120 & 4.76 \\
    \hline
    2 & 0.024 & 4 & 0.25 & 25 & 8.565 & 120 & 4.76 \\
    \hline
    3 & 0.024 & 5 & 0.25 & 25 & 4.226 & 120 & 4.76 \\
    \hline
    4 & 0.024 & 6 & 0.25 & 25 & 9.430 & 120 & 4.76 \\
    \hline
    5 & 0.024 & 7 & 0.25 & 25 & 8.353 & 120 & 4.76 \\
    \hline
    6 & 0.024 & 5 & 0.20 & 25 & 4.686 & 120 & 4.76 \\
    \hline
    7 & 0.024 & 5 & 0.30 & 25 & 3.873 & 120 & 4.76 \\
    \hline
    8 & 0.024 & 5 & 0.35 & 25 & 7.417 & 120 & 4.76 \\
    \hline
    9 & 0.024 & 5 & 0.40 & 25 & 3.600 & 120 & 4.76 \\
    \hline
    10 & 0.024 & 5 & 0.45 & 25 & 2.602 & 130 & 5.16 \\
    \hline
    11 & 0.024 & 5 & 0.25 & 30 & 4.021 & 133 & 4.4 \\
    \hline
    12 & 0.024 & 5 & 0.25 & 35 & 9.147 & 133 & 3.77 \\
    \hline
    13 & 0.024 & 5 & 0.25 & 40 & 8.153 & 148 & 3.68 \\
    \hline
    14 & 0.024 & 5 & 0.25 & 45 & 9.197 & 151 & 3.33 \\
    \hline
    15 & 0.024 & 5 & 0.25 & 50 & 5.023 & 168 & 3.34 \\
    \hline
    16 & 0.021 & 5 & 0.25 & 25 & 6.106 & 147 & 5.84 \\
    \hline
    17 & 0.027 & 5 & 0.25 & 25 & 7.191 & 111 & 4.40 \\
    \hline
    18 & 0.030 & 5 & 0.25 & 25 & 8.231 & 103 & 4.08 \\
    \hline
    19 & 0.033 & 5 & 0.25 & 25 & 4.517 & 99 & 3.92 \\
    \hline
\end{tabular}

\includegraphics[width=480pt]{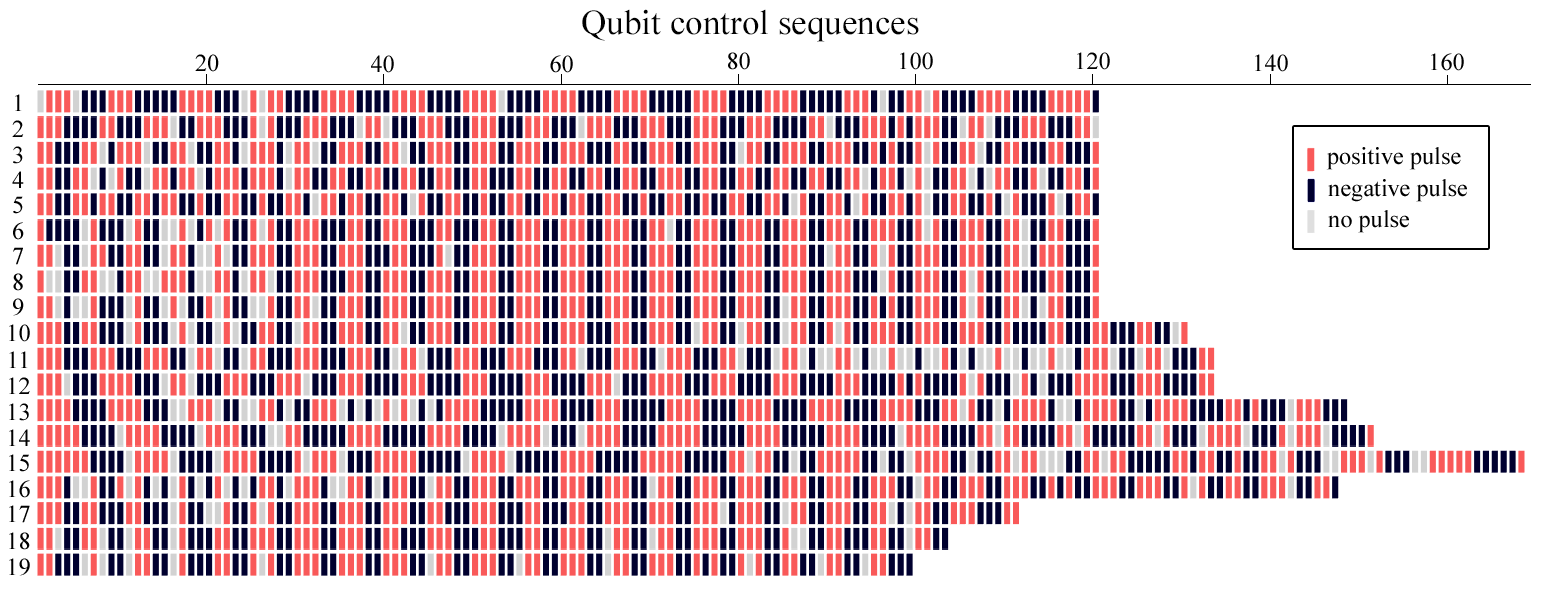}

\end{table*}

\section{Control sequence search results and discussion}

Let's now find the optimal bipolar control sequences for a set of typical system parameters. We present the parameter values, $\Delta\theta$, $\omega_{01}/2\pi$, $\mu/2\pi$, $\omega_g/2\pi$, as well as gate infidelity, $1 - \langle F \rangle$, sequence length, and gate time, $t_{\pi/2}$, in Table 1. The sequences themselves are shown under the table.

\begin{figure}[t]
\center{\includegraphics[scale=0.9]{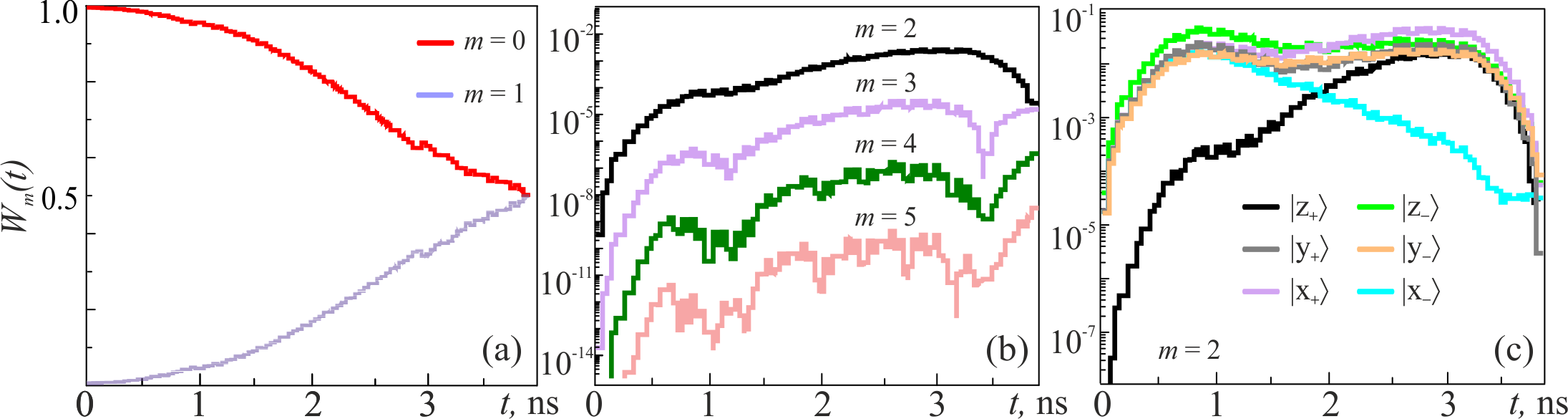}}
\caption{Evolution of the qubit state populations under the drive with control sequence No.19 (see the system parameters in Table~1). (a) and (b) the basic qubit levels and the upper levels, accordingly; the qubit is initialized in the state $|z_{+}\rangle$. (c) The leakage to the second excited level for the qubit initialized at different poles on the Bloch sphere.}
\label{Fig4:image}
\end{figure}

An example of the dynamics of the basic qubit state populations, $W_{0,1}(t)$, for sequence No.19 is shown in figure~\ref{Fig4:image}a. The qubit is initialized in the state $|z_{+}\rangle$. The system parameters are close to the ones considered in Ref.~\cite{McD1}. Figure~\ref{Fig4:image}b shows that the leakage to the state $|2\rangle$ is the greatest among the states outside the computational basis ($m=2,3,4,5$). Figure~\ref{Fig4:image}c presents the leakage to the second excited level for the qubit initialized in the different poles on the Bloch sphere, defined by expressions (\ref{eq5}). While the leakage value may be greater than $10^{-2}$ during the operation, it is less than $10^{-4}$ at the end of the gate. 
Note that the gate time provided by the bipolar drive, $t_{\pi/2} = 3.92$~ns, is less than half that obtained with the sequences \cite{McD1} at similar gate fidelity.

\begin{figure}
\center{\includegraphics[scale=0.95]{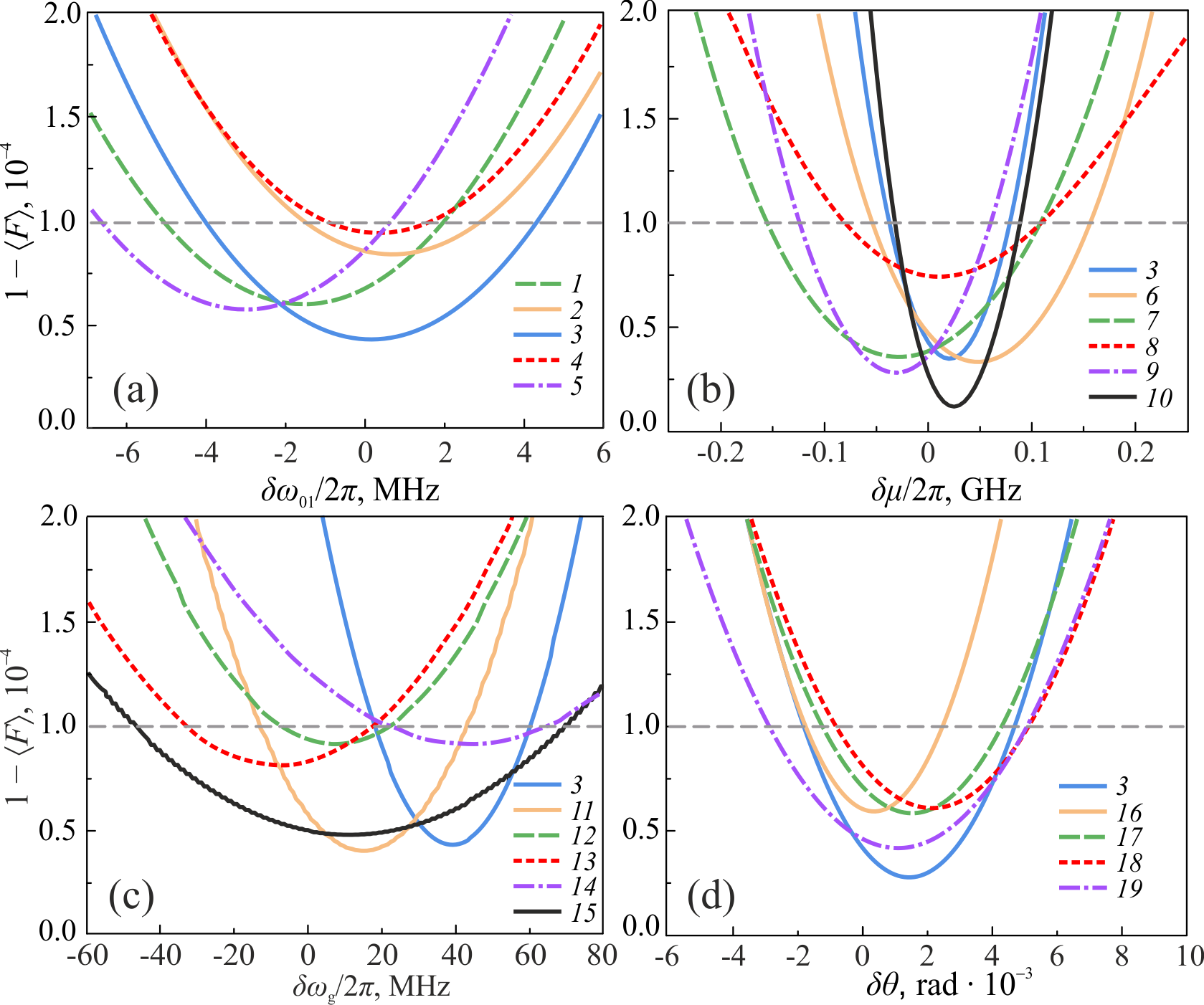}}
\caption{$Y_{\pi/2}$ gate infidelity, $1 - \langle F \rangle$, versus detuning of the system parameters: (a) qubit frequency, $\delta\omega_{01}$; (b) anharmonicity, $\delta\mu$; (c) clock generator frequency, $\delta\omega_g$; (d) single pulse rotation angle, $\delta\theta$. The control sequence numbers are shown in legends.}
\label{Fig5:image}
\end{figure}

We investigate the gate robustness in variations of the system parameters in terms of stability of the fidelity. To do this, we introduce a detuning from the found optimal values: $\Delta\theta \pm \delta\theta$, $\omega_{01} \pm \delta\omega_{01}$, $\mu \pm \delta\mu$, $\omega_g \pm \delta\omega_g$. Figure~\ref{Fig5:image} shows the change in the infidelity depending on the introduced detuning for all rows of the Table~1. The infidelity value $1 - \langle F \rangle = 10^{-4}$ is presented as a guide for eye by horizontal dashed line.

We find that while the qubit frequency shift $\delta\omega_{01}/2\pi$ can be of several MHz to keep the fidelity over 99.99\%, the same shift in nonlinearity, $\delta\mu/2\pi$, may be almost two orders of magnitude larger, see figure~\ref{Fig5:image}a,b. The gate fidelity is also weakly dependent on the change in the frequency of the clock generator, which can be shifted by tens of MHz, figure~\ref{Fig5:image}c. Our analysis shows that the most influential parameter is the amplitude of the SFQ pulse coupled to the qubit. Here it is represented by a single pulse rotation angle (tolerance $\delta\theta \sim 0.005$~rad, figure~\ref{Fig5:image}d). Therefore, its value must be well estimated for the appropriate control sequence search. The results presented in figure~\ref{Fig5:image} confirm that the proposed optimization algorithm allows one to find a bipolar SFQ control sequence for arbitrary parameter values from the practical range to execute a single qubit gate with fidelity of more than 99.99\%. Note that due to the discrete nature of the driving SFQ pulse signal, the minima of the curves shown in figure~\ref{Fig5:image} don't correspond to zero value of parameter variation.

The found sequences may be pretty long (more than 150 symbols), especially in cases of high generator frequencies. These sequences could be divided into repeated subsequences to save memory \cite{QCIMukh,McD1}. It is seen that all the found sequences presented beneath the Table~1 have recurrent patterns. A deeper study of the symmetries of the solutions is desirable to determine whether they are associated with any particular multiplicity of clock and qubit frequencies, and whether they can be found for arbitrary system parameters. 

There are several directions for further study. We have presented the simplest way to obtain a regular bipolar SFQ pulse signal using differential connections of two DC-to-SFQ converters. The utilization of the considered tritstreams is possible by using the existed binary superconducting digital circuits, though implies hardware overhead. Designing circuit solutions at the level of ternary logic cells, as well as the entire device architecture, can improve the performance of an SFQ co-processor.

Finally, the proposed optimization algorithm may be improved itself. The genetic algorithm could be used to enlarge the number of found solutions in order to examine the problem landscape, studying the properties of local minima, and finding an approach to get to the global one. Latest researches also show that the utilization of deep learning is an effective way, since it works well with unipolar SCALLOP sequences \cite{alphazero}.

\section{Conclusion}

We consider the possible application of bipolar SFQ control for the implementation of a single qubit gate. The robust optimization algorithm for SFQ control sequence search is presented. The algorithm has been demonstrated to find an appropriate sequence for a $\pi / 2$ Y-rotation for arbitrary system parameters from a practical range: $\omega_{01}/2\pi \in [3...7]$~GHz, $\mu/2\pi \in [0.2...0.45]$~GHz, the single pulse rotation angle, $\Delta\theta \in [0.021...0.033]$~rad, and the clock generator frequency, $\omega_g/2\pi \in [25...50]$~GHz. The gate length provided by the found sequences is approximately two times less than that under the unipolar SCALLOPS proposed earlier, and hence it is no slower than under conventional microwave drive. The ranges of the system parameter deviations where the gate fidelity remains above 99.99\% are suitable for practical applications. The directions for further development of the proposed bipolar SFQ control is briefly discussed.

\section*{Acknowledgments}
The development of the optimization algorithm is performed under support of the Grant No.~20-69-47013 of the Russian Science Foundation. The numerical simulations were supported by UNN within the framework of the strategic academic leadership program ``Priority 2030" of Ministry of Science and Higher Education of the Russian Federation. V.V. is grateful to the Interdisciplinary Scientific and Educational School of Moscow State University ``Photonic and Quantum Technologies. Digital Medicine". The work of A.S. on the section ``Transmon under SFQ drive" was carried out with the support of the RSF project No.~22-21-00586. Thanks to the Basis Foundation scholarship for supporting the work.

\section*{References}
\bibliographystyle{iopart-num}
\bibliography{bibliography}

\end{document}


\title{Supplementary materials for ``Speeding up qubit control with bipolar single-flux-quantum pulse sequences"}

\author{Vsevolod Vozhakov$^{1,2,3}$, Marina Bastrakova$^{2}$, Nikolay Klenov$^{2,4}$, Arkady Satanin$^{5}$, Igor Soloviev$^{1,2}$}
\address{$^1$Lomonosov Moscow State University Skobeltsyn Institute of Nuclear Physics, 119991 Moscow, Russia}
\address{$^2$Lobachevsky State University of Nizhny Novgorod, Nizhny Novgorod 603950, Russia}
\address{$^3$Russian Quantum Center, 143025 Skolkovo, Moscow, Russia}
\address{$^4$Faculty of Physics, Lomonosov Moscow State University, 119991 Moscow, Russia}
\address{$^5$Russia National Research University Higher School of Economics, Moscow 101000, Russia}

\ead{igor.soloviev@gmail.com}

\section{Comparison of different shapes of pulses in a regular drive sequence}

In our calculations, we have replaced the real shape of an SFQ voltage pulse with a rectangular one for the reasons described in the main text. Here we present an example of $\pi / 2$ rotation of a transmon with frequency, $\omega_{01} = 5$~GHz, and aharmonicity, $\mu = 250$~MGz, under three pulse driving signals: with pulses having rectangular, Gauss and actual SFQ voltage pulse shape numerically calculated on the output Josephson junction (with typical characteristic frequency, $\omega_c = 250$~GHz) of DC-to-SFQ converter \cite{likharev1991} using PSCAN software \cite{pscan}. The pulse width was chosen so that their time integrals were the same. The pulse repetition rate in all cases corresponds to the qubit frequency, $\omega_{01}$. 

\begin{figure}[h]
\center{\includegraphics[scale=0.25]{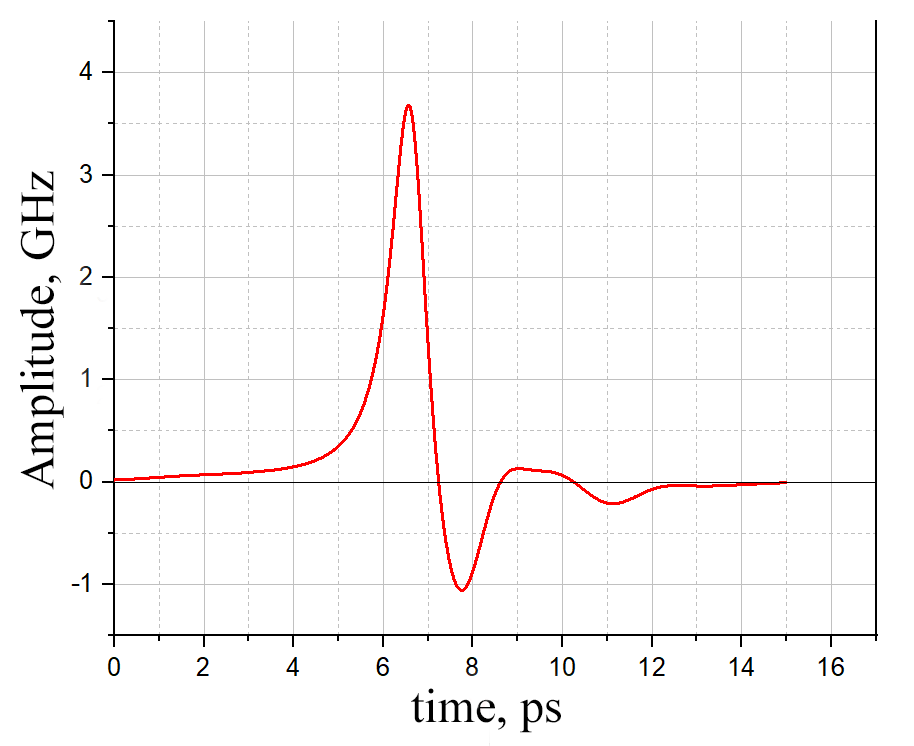}}
\caption{Shape of an SFQ pulse numerically calculated on the output Josephson junction of DC-to-SFQ converter.}
\label{algscheme:image1}
\end{figure}

\begin{figure}[h]
\center{\includegraphics[scale=0.45]{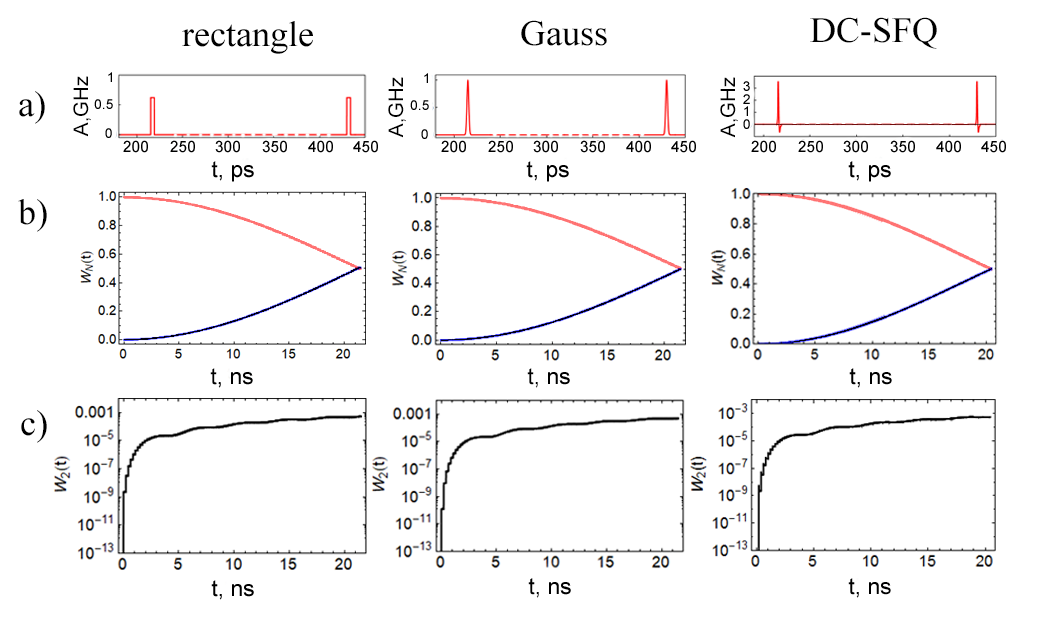}}
\caption{$\pi / 2$ rotation of a transmon with frequency, $\omega_{01} = 5$~GHz, and aharmonicity, $\mu = 250$~MGz, under three pulse driving signals: with pulses having rectangular, Gauss and SFQ voltage pulse shape (see figure~\ref{algscheme:image1}). a) Drive signals with amplitude of pulses aligned to perform the chosen quantum gate. b) Qubit evolution, \textit{red line} -- ground state probability, \textit{blue line} -- excited state probability. c) Leakage to the first excited state outside the computational basis.}
\label{algscheme:image2}
\end{figure}

The calculated SFQ pulse shape is shown in figure~\ref{algscheme:image1}. The pulses shape in driving signals is presented in figure~\ref{algscheme:image2}, panels (a). The amplitude of these pulses was aligned to perform the chosen quantum gate. The panels (b) and (c) of figure~\ref{algscheme:image2} show the evolution of population of the two first qubit levels and the one of the first excited level outside the computational basis, correspondingly. The obtained consistency of the numerical simulation results illustrates the validity of our approximation for SFQ pulse shape.


Another important point to consider here is the width of the pulse. In our simulations, we use a rectangular pulse with a width equal to the time grid step (0.01 ps) to numerically solve the Schrodinger equation. This allows to treat the pulse almost as a delta-function that makes our results comparable to the ones obtained in the previous works \cite{McD0,McD1}. However, it was shown \cite{McD2023} that there may be a need for the pulse broadening in the real qubit digital control system, which may outflows in some gate fidelity degradation \cite{McD0}.

\section{Final state wave function phase}
One can see that right after the application of sequences found using the optimization algorithm described in Section 5 of the main text, the phase of the resulting wave function is different from that obtained by the implementation of an ideal $Y_{\pi/2}$ operation. This phase difference can be effectively eliminated by the free qubit precession in an appropriate time less than $2\pi/\omega_{01}$. Another approach is a ``virtual Z-gate'' \cite{zgate} with no time penalty, nor additional leakage outside the computational basis. Therefore, all calculations were implemented without accounting for this phase to expand the search space.

\section*{References}
\bibliographystyle{iopart-num}
\bibliography{bibliography}